\documentclass[conference]{IEEEtran} 
\IEEEoverridecommandlockouts
\usepackage{amsmath,graphicx, amsfonts, dsfont, stmaryrd, svg, url, amssymb}
\usepackage[bookmarks,colorlinks]{hyperref}
\usepackage{acronym, cite}
\usepackage{colortbl} 
\usepackage{relsize}
\usepackage{siunitx}
\usepackage{booktabs}
\usepackage{fontawesome5}

\usepackage{siunitx}
\usepackage[ruled,linesnumbered,vlined]{algorithm2e}
\let\oldnl\nl
\newcommand{\nonl}{\renewcommand{\nl}{\let\nl\oldnl}}
 
\usepackage{enumitem}

\newcommand{\myVec}[1]{{\boldsymbol{#1}}}
\newcommand{\myMat}[1]{{\boldsymbol{#1}}}
\newcommand{\mySet}[1]{\mathcal{#1}}

\newtheorem{example}{Example}

\acrodef{ai}[AI]{artificial intelligence}
\acrodef{dnn}[DNN]{deep neural network}
\acrodef{sgd}[SGD]{stochastic gradient descent}
\acrodef{mimo}[MIMO]{multiple-input multiple-output}
\acrodef{sic}[SIC]{soft interference cancellation}
\acrodef{mlp}[MLP]{multi-layer perceptron}
\acrodef{ekf}[EKF]{extended Kalman filter}
\acrodef{vi}[VI]{variational inference}
\acrodef{llr}[LLR]{log-likelihood ratio}
\acrodef{ofdm}[OFDM]{orthogonal frequency division multiplexing}
\acrodef{lmmse}[LMMSE]{linear minimum mean-squared error}
\acrodef{zf}[ZF]{zero-forcing}
\acrodef{cfo}[CFO]{carrier frequency offset}
\acrodef{iqmm}[IQMM]{IQ mismatch}
\acrodef{ici}[ICI]{inter-carrier interference}
\acrodef{cnn}[CNN]{convolutional neural network}
\acrodef{ssm}[SSM]{state space model}
\acrodef{snr}[SNR]{signal-to-noise ratio}
\acrodef{map}[MAP]{maximum {\em{a-posteriori}}}
\acrodef{ourdnn}[ESCNN]{element-wise scaled \ac{cnn}}

 \usepackage[all=normal,paragraphs=tight,floats=normal,mathspacing=normal,wordspacing=tight,charwidths=tight,mathdisplays=normal,leading=normal]{savetrees}

\begin{document}

\title{Neural Augmentation of MIMO-OFDM Receivers for Universal LLR  Reconstruction
}

\author{Ory Eger,~\IEEEmembership{{Student Member},~IEEE}, and Nir Shlezinger,~\IEEEmembership{Senior Member,~IEEE}
\thanks{
Parts of this work were presented at the 2026 IEEE International Conference on Communications (ICC) as the paper~\cite{eger2026learning}.
 The authors are with the School of ECE,  Ben-Gurion University of the Negev, Be’er-Sheva, Israel (e-mail: egero@post.bgu.ac.il; nirshl@bgu.ac.il).  The work was supported by the European Research Council (ERC) under the ERC starting grant nr. 101163973 (FLAIR).}}

\maketitle


\begin{abstract}
The growing demands for higher throughput and cost-efficient wireless communications drive the need for  receivers that are both simple to deploy and robust to hardware impairments and nonlinear environments. While classical model-based receivers and recently proposed \ac{dnn} architectures provide complementary benefits, they either rely on simplified linear Gaussian assumptions, require considerable computational resources, or are tailored for a given setting and modulation. In this work, we propose a compact and modular \ac{dnn} augmentation that universally refines the soft outputs of existing receivers (model-based or data-driven), addressing two distinct operating regimes: structurally incomplete soft information arising from reduced-complexity detectors, and degraded soft outputs caused by hardware impairments and synchronization errors. A key property of the proposed framework is its task-agnostic nature: operating without any knowledge of the specific source of unreliability, it produces well-calibrated \acp{llr} suitable for channel decoding. Our design leverages an element-wise scaled convolutional neural network tailored to perform learned interference cancellation across users and neighboring subcarriers, combined with a training algorithm that encourages accurate \acp{llr} for soft channel decoding. Numerical results demonstrate that the proposed augmentation consistently improves diverse receiver algorithms in challenging channel conditions while incurring minimal overhead.
\end{abstract}

\acresetall

\section{Introduction}
\label{sec:intro}
Wireless communication systems are facing rapidly growing demands in terms of throughput, reliability,  cost, and energy-efficiency~\cite{wang2023road}. To meet these requirements, modern networks increasingly rely on multi-user \ac{mimo} architectures combined with multi-carrier waveforms such as \ac{ofdm}~\cite{saad2019vision}. While these technologies enable high spectral efficiency and flexible resource allocation, they also impose stringent requirements on receiver design. In particular, practical receivers are expected to operate under tight computational and latency constraints, while coping with challenging propagation conditions, multi-user interference, and various hardware-induced non-idealities. These competing requirements make it difficult to simultaneously achieve high detection performance, robustness to mismatches, and low implementation complexity.


In multi-user \ac{mimo}-\ac{ofdm} systems, receiver processing has been traditionally dominated by classical model-based approaches~\cite{albreem2019massive}. These include low-complexity linear receivers such as \ac{zf} and \ac{lmmse} equalizers~\cite{yang2015fifty}, as well as more advanced algorithms such as sphere decoding~\cite{hung2006sphere} and soft interference cancellation~\cite{choi2000}. While these methods are well understood, analytically grounded, and widely deployed, they fundamentally rely on the assumption that the underlying channel can be accurately described as a \emph{linear (pseudo) time-invariant system with additive Gaussian noise}. In practice, however, modern wireless systems are often subject to hardware mismatches and impairments, such as \ac{iqmm} and amplifier nonlinearities, as well as \ac{cfo}, which in multi-carrier settings further induces \ac{ici}~\cite{mohammadian2021rf}. These mismatches all affect the classic canonical model. As a result, the performance of conventional receivers can significantly degrade when operating outside their assumed regime. Addressing these effects within the model-based paradigm typically requires either highly specialized compensation schemes tailored to specific impairments (e.g., low-resolution quantization~\cite{li2017channel, shlezinger2019asymptotic}), iterative interference cancellation methods for \ac{ici} mitigation~\cite{molisch2007iterative}, or the adoption of more general but computationally demanding models (e.g., Volterra series expansions~\cite{postcompensation2011}). Consequently, such approaches either lack robustness and universality to diverse real-world mismatches or incur notable complexity, limiting their suitability for next-generation wireless systems.

Recent advances in \ac{ai} have led to the emergence of a broad family of \acp{dnn} for receiver processing~\cite{dai2020deep}. Approaches based on recurrent neural networks~\cite{farsad2018neural}, \acp{cnn}~\cite{goutay2021machine,farhadi2023deep}, residual networks~\cite{honkala2021deeprx,pihlajasalo2023deep,pihlajasalo2025multidilated}, graph neural networks~\cite{clausius2025joint,zhou2023graph,li2025hypergepnet, cammerer2023neural}, variational autoencoders~\cite{caciularu2020unsupervised}, and attention mechanisms~\cite{xie2023comm,michon2022convolutional} have been proposed to replace conventional receiver modules (see survey in~\cite{doha2025deep}). 
By learning the input--output mapping directly from data, these architectures can operate effectively in complex and poorly modeled environments without requiring explicit channel knowledge. However, such black-box designs are typically characterized by high model complexity, substantial training data requirements, and limited interpretability.
Moreover, deploying such architectures in practice raises significant challenges related to real-time inference~\cite{wiesmayr2025design},  as well as dynamic reconfiguration (e.g., to varying modulation orders) and site-specific adaptation~\cite{raviv2024adaptive}.

An alternative line of work, rooted in model-based deep learning~\cite{shlezinger2023model}, aims to bridge the gap between classical signal processing and data-driven methods by embedding domain knowledge into the network architecture. This includes approaches that unfold iterative algorithms into trainable layers~\cite{deka2026comprehensive,balatsoukas2019deep,shlezinger2025deep}, augment conventional receivers with \acp{dnn}-based learned components~\cite{sun2025comprehensive,honkala2026eqdeeprx,bai2025deep,fesl2026learning}, or replace complex computations such as symbol-to-\ac{llr} mappings~\cite{shental2019machine}. Such hybrid designs typically offer improved interpretability, reduced parameterization, and faster training compared to fully data-driven \acp{dnn}. Nevertheless, existing methods often inherit the limitations of their underlying model-based formulations (e.g., the reliance on linear Gaussian channel modeling)~\cite{khani2020adaptive, he2018model,liu2025gnn}. Channel model-agnostic hybrid architectures are often  tailored to simplified settings such as narrowband systems~\cite{DeepSIC2021,van2022deep,loli2023model}, and extending these approaches to multi-carrier scenarios by independently processing each subcarrier leads to increased computational complexity and fails to capture cross-subband effects, such as \ac{ici}.

\subsection*{Main Contributions}
\begin{figure}
    \centering
    \includegraphics[width=\linewidth]{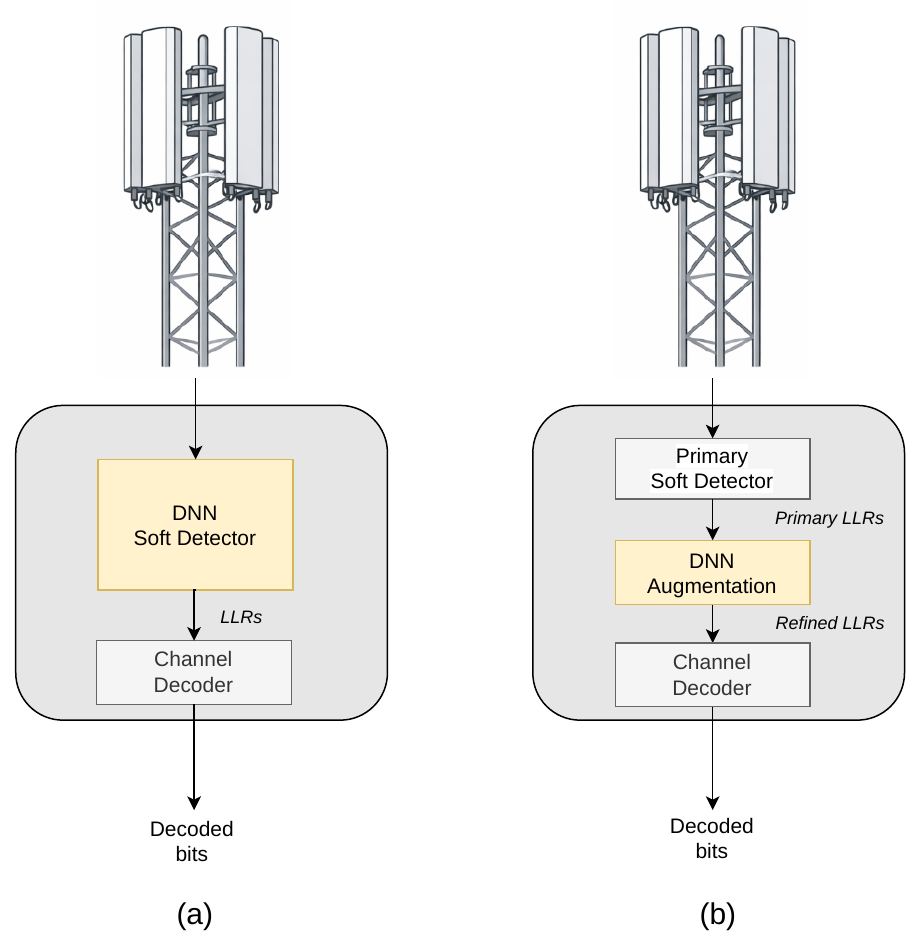}
    \vspace{-0.4cm}
    \caption{DNN for soft detection: $(a)$ A DNN serves as an end-to-end soft detector, directly producing LLRs for the channel decoder; $(b)$ The proposed approach: A DNN augments some primary soft detector, refining its output LLRs before channel decoding}
    \label{fig:e2e_vs_Augment}
    \vspace{-0.2cm}
\end{figure}

The diversity of receiver algorithms, each offering a different trade-off between complexity, robustness, and performance, motivates a paradigm shift from designing standalone receivers toward developing {\em augmentative} mechanisms that enhance existing processing chains. In this work, we propose a lightweight neural augmentation framework for multi-user \ac{mimo}-\ac{ofdm} receivers, that operates by refining the soft outputs of an arbitrary primary detector, which can be model-based, data-driven, or hybrid. By focusing on {\em post-processing augmentation} rather than end-to-end replacement as shown in Fig.~\ref{fig:e2e_vs_Augment}, the proposed approach preserves the structure and reliability of the underlying receiver while significantly improving its robustness to model mismatches (including \ac{ici}, synchronization errors, or hardware-induced nonlinearities), and overcome specificity to a given modulation. This design enables a compact, modular, and scalable solution that can be seamlessly integrated into existing systems, providing a unified mechanism for enhancing diverse receiver architectures and extending operation regimes under challenging operating conditions.

Our main contributions are summarized as follows:
\begin{itemize}
    \item \textbf{Modular neural augmentation for soft detection:}
    We propose a general framework that augments arbitrary MIMO-OFDM receivers by refining their bit-wise \acp{llr}. The proposed module is agnostic to the structure of the primary detector, and can be applied on top of model-based, deep learning-based, and hybrid receivers.

    \item \textbf{Lightweight architecture for cross-subcarrier refinement:}
    We develop our compact augmentation network based on an  \ac{ourdnn}, which exploits local spectral correlations to perform learned interference mitigation across neighboring subcarriers and users. The architecture is highly parameter-efficient and tailored for low-latency implementation.

    \item \textbf{Robust and scalable enhancement under mismatches:}
    The proposed augmentation effectively compensates for a wide range of impairments, including \ac{ici}, \ac{cfo}, \ac{iqmm} and clipping distortion, enabling conventional receivers to operate reliably beyond their nominal assumptions. It further supports heterogeneous receiver configurations, including high-complexity detectors such as sphere decoding, and enables reduced-complexity operation via \ac{llr} completion, where only a subset of bits is estimated and the remaining ones are reconstructed by the augmentation module.

    \item \textbf{Flexible training and modulation-agnostic design:} We devise a training methodology that explicitly promotes accurate probabilistic outputs, yielding well-calibrated \acp{llr} suitable for channel decoding. Combined with the bit-wise formulation, this allows a single trained augmentation module to support multiple modulation orders, facilitating scalable deployment across varying system configurations. This further enables a unified treatment of two distinct operating regimes: \ac{llr} completion, where the primary detector provides only a subset of bit \acp{llr}, and impairment-induced \ac{llr} corruption, where hardware or channel non-idealities distort the soft outputs — without requiring knowledge of the operating regime or the nature of the impairments.

    \item \textbf{Comprehensive numerical validation:}
    Through simulations with physically compliant channels based on 3GPP specifications~\cite{3gpp_tr38901} and modeled using Sionna~\cite{sionna2022} and Quadriga \cite{jaeckel2014quadriga}, we demonstrate that the proposed augmentation consistently improves the performance of diverse receivers under challenging conditions, while incurring negligible computational and latency overhead. 
\end{itemize}

\subsection*{Organization and Notations}

The rest of this paper is structured as follows. Section~\ref{sec:System Model and Preliminaries} presents the system model. Section~\ref{sec: Method} details the proposed augmentation, which we evaluate in  Section~\ref{sec: Numerical Study}. Section~\ref{sec:conclusion} provides concluding remarks.

Throughout this paper, we use boldface lowercase and uppercase letters for vectors (e.g. $\myVec{x}$) and matrices (e.g. $\myMat{M}$), respectively. The $i$th element of  $\myVec{x}$ is denoted $[\myVec{x}]_i$, $[\myMat{M}]_{i,j}$ is the $(i,j)$th element of a matrix $\myMat{M}$, while $[\myMat{M}]_{i,:}$ is its $i$th row.
We use $\myMat{I}$ for the identity matrix, $\mathbb{R}$ for the set of real numbers, and $\mathbb{C}$ for the set of complex numbers. 


\section{System Model and Preliminaries}
\label{sec:System Model and Preliminaries}  

\subsection{System Model}
\label{subsec: Signal Model}
We consider a multi-user multi-carrier uplink \ac{mimo} communication system, where $K$ single-antenna users transmit to a base station equipped with $n_r$ receive antennas over $B$ orthogonal subcarriers. Transmission is organized in discrete time blocks indexed by $i$, where each block corresponds to a single \ac{ofdm} symbol in the frequency domain.

{\bf Channel Input:}
Let $\myMat{S}^{\,i} \in \mySet{S}^{B \times K}$ denote the matrix of transmitted symbols in the $i$th block, where $[\myMat{S}^{\,i}]_{b,k}$ is the complex symbol transmitted by user $k$ over subcarrier $b$. The symbol alphabet $\mySet{S} \subset \mathbb{C}$ represents a digital modulation constellation of size $|\mySet{S}| = 2^{N_s}$, where each symbol encodes $N_s$ information bits (e.g., $N_s=4$ for 16-QAM). The corresponding transmitted bits are represented by the tensor $\myMat{D}^i \in \{0,1\}^{B \times K \times N_s}$, where $d^i_{b,k,n} = [\myMat{D}^i]_{b,k,n}$ denotes the $n$th bit associated with the symbol transmitted by user $k$ on subcarrier $b$ at block $i$. This representation explicitly captures the bit-wise structure required for soft detection and subsequent channel decoding.

\begin{figure}
    \centering
    \includegraphics[width=\linewidth]{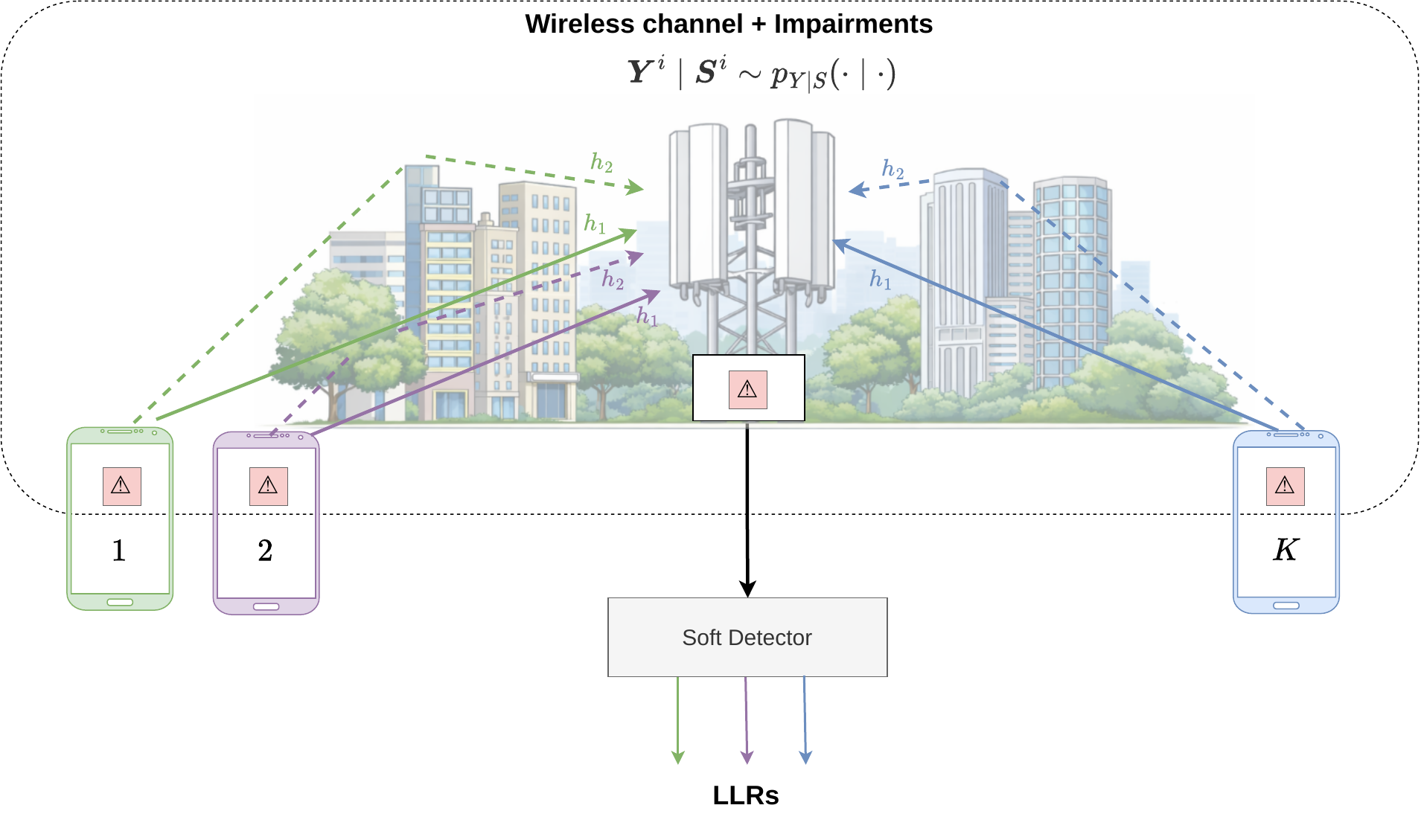}
    \vspace{-0.4cm}
    \caption{System model for $K$ users. Pink blocks denote hardware impairments arising in the RF front-end and digital front-end at both transmitter and receiver}
    \label{fig:System}
    \vspace{-0.2cm}
\end{figure}

{\bf Channel Output:}
The received signal is given by the matrix $\myMat{Y}^i \in \mathbb{C}^{B \times n_r}$, where $[\myMat{Y}^i]_{b,:}$ denotes the $n_r$-dimensional received vector corresponding to subcarrier $b$. Rather than committing to a specific parametric model, we adopt a general stochastic formulation in which the input-output relation is characterized by the conditional distribution $p_{\myMat{Y^i}|\myMat{S^i}}(\cdot \mid \cdot)$,
that captures the combined effect of the wireless propagation channel, multi-user interference, and the full transmit and receive chains, including synchronization mismatches and hardware impairments (see Fig.~\ref{fig:System}).
We assume a stationary setting in which the conditional distribution does not depend on the block index $i$, i.e., 
\begin{equation}
\myMat{Y}^i \mid \myMat{S}^{\,i} \sim p_{\myMat{Y}|\myMat{S}}(\cdot \mid \cdot), \qquad  \forall i \in \{1,2,\ldots\}. 
\end{equation} 


 
While the formulation above is intentionally general, it encompasses classical channel models as well as systems affected by hardware impairments. We show this with two examples, beginning with the linear Gaussian channel model and followed by an OFDM system with \ac{cfo}.

\begin{example}
    \label{exm:LinGauss}
    In linear Gaussian channels, the received signal on each subcarrier is modeled as a linear transformation of the transmitted symbols corrupted by additive Gaussian noise. Specifically, for each subcarrier $b \in \{1,\ldots,B\}$, the received vector for the $i$th OFDM symbol $\myVec{y}_b^i \in \mathbb{C}^{n_r}$ is given by
    \begin{equation}
    \myVec{y}_b^i = \myMat{H}_b \myVec{s}_b^i + \myVec{w}_b^i,
    \end{equation}
    where $\myVec{s}_b^i \in \mathbb{C}^{K}$ collects the symbols transmitted by all $K$ users on subcarrier $b$, $\myMat{H}_b \in \mathbb{C}^{n_r \times K}$ is the corresponding channel matrix, and $\myVec{w}_b^i $ is additive circularly symmetric complex Gaussian noise with variance $\rho^2$ per dimension.

    Under the additional assumption that different subcarriers are statistically independent, the conditional distribution factorizes as
    \begin{equation*}
        p_{\myMat{Y}|\myMat{S}}\!\left([\myVec{y}_1^i,\ldots,\myVec{y}_B^i]\middle|[\myVec{s}_1^i,\ldots,\myVec{s}_B^i]\right)
        = \prod_{b=1}^B \mathcal{CN}\!\left(\myVec{y}_b^i; \myMat{H}_b \myVec{s}_b^i, \rho^2 \myMat{I}\right),
    \end{equation*}
    where $\mathcal{CN}(\cdot;\myVec{\mu},\myMat{\Sigma})$ denotes the multivariate complex Gaussian distribution with mean $\myVec{\mu}$ and covariance $\myMat{\Sigma}$.
\end{example}
\begin{example}
    \label{exm:CFO_ICI}
    In OFDM systems affected by a \ac{cfo} of $\varepsilon$ (normalized to the subcarrier spacing), the received signal on each subcarrier is modeled as a linear combination of the transmitted symbols on \emph{all} subcarriers,
    corrupted by additive Gaussian noise~\cite{shaked2017joint}. Specifically, for each subcarrier
    $b \in \{1,\ldots,B\}$ at OFDM symbol $m$, the received vector
    $\myVec{y}_b^i \in \mathbb{C}^{n_r}$ is given by
    \begin{equation}
        \myVec{y}_b^i = e^{j\phi_i}
        \sum_{l=1}^{B} c_{l,b}(\varepsilon)\,
        \myMat{H}_l  \myVec{s}_l^i + \myVec{w}_b^i,
    \end{equation}
    where $\myVec{s}_l^i \in \mathbb{C}^K$, $\myMat{H}_b \in \mathbb{C}^{n_r \times K}$,
    and $\myVec{w}_b^{i}$ are as in
    Example~\ref{exm:LinGauss}, $\phi_i$ is a symbol-dependent phase rotation induced
    by the CFO, and $c_{l,b}(\varepsilon)$ is the \ac{ici}
    coefficient describing the leakage from subcarrier $l$ onto subcarrier $b$, with $c_{l,b}(\varepsilon) \to 0$
    for $b \neq l$ as $\varepsilon \to 0$.
    Since the noise is independent across subcarriers, the conditional distribution
    factorizes as
    \begin{multline*}
        p_{\myMat{Y}|\myMat{S}}\!\left(
            [\myVec{y}_1^{i},\ldots,\myVec{y}_B^{i}]
            \middle|
            [\myVec{s}_1^{i},\ldots,\myVec{s}_B^{i}]
        \right) \\
        = \prod_{b=1}^{B}
        \mathcal{CN}\!\left(
            \myVec{y}_b^{i};\,
            e^{j\phi_i}\!\sum_{l=1}^{B} c_{l,b}(\varepsilon)\,
            \myMat{H}_l \myVec{s}_l^{i},\,
            \rho^2 \myMat{I}
        \right).
    \end{multline*}
    Unlike Example~\ref{exm:LinGauss}, the received signal on each subcarrier
    depends on the transmitted symbols across \emph{all} $B$ subcarriers rather than on a single subcarrier alone.    
\end{example}


Beyond channel-induced distortions, in reduced-complexity receivers, computational constraints may limit the detector to recovering only a subset of the $N_s$ bits per symbol, leaving the remaining bits undetected. While the channel itself may be free of impairments, the receiver architecture inherently provides incomplete soft information as a deliberate trade-off between detection quality and computational complexity. We further elaborate on such receiver-induced distortions in the sequel.



\subsection{Soft Symbol Detection}
\label{ssec:Detection}
We focus on the {\em soft detection} receiver task, whose goal is to map the received signal $\myMat{Y}^i$ into probabilistic estimates of the transmitted bits. In particular, we aim to recover the bit-wise \acp{llr}, denoted $\{l_{b,k,n}^i\}$, which quantify the reliability of each bit and serve as the input to channel decoders. These are defined as
\begin{equation}
    \label{eqn:LLR}
    l_{b,k,n}^i = \log \frac{\Pr(d^i_{b,k,n} = 1 \mid \myMat{Y}^i)}{\Pr(d^i_{b,k,n} = 0 \mid \myMat{Y}^i)}.
\end{equation}
Accurate estimation of these \acp{llr} is critical for reliable decoding, as they directly determine the performance of subsequent forward error correction schemes.

A wide range of receiver algorithms can be interpreted as mappings from the channel observations $\myMat{Y}^i$ to estimates of the \acp{llr} in~\eqref{eqn:LLR}. Classical model-based approaches, typically derived under the linear Gaussian assumption of Example~\ref{exm:LinGauss}, compute approximate \acp{llr} using linear equalizers such as \ac{zf} and \ac{lmmse}~\cite{yang2015fifty}, or more sophisticated detectors such as sphere decoding~\cite{hung2006sphere} and soft interference cancellation~\cite{choi2000}. These methods leverage analytical models of the channel to derive tractable approximations of the posterior probabilities.

More recently, data-driven approaches based on deep learning have emerged, including fully end-to-end \acp{dnn}~\cite{honkala2021deeprx} as well as hybrid model-based/data-driven architectures~\cite{DeepSIC2021}. Such methods learn to approximate the mapping $\myMat{Y}^i \mapsto \{l_{b,k,n}^i\}$ directly from data, and can therefore operate effectively even when the underlying channel deviates from standard modeling assumptions. However, many existing approaches remain tailored to simplified settings, such as narrowband systems, effectively assuming independence across subcarriers, i.e.,
\begin{equation*}
    p_{\myMat{Y}|\myMat{S}}\left([\myVec{y}_1^{i},\ldots,\myVec{y}_B^{i}]\middle|[\myVec{s}_1^{i},\ldots,\myVec{s}_B^{i}]\right) = \prod_{b=1}^B p_{[\myMat{Y}]_{b,:}\mid[\myMat{S}]_{b,:}}\left([\myVec{y}_b^{i}\middle|[\myVec{s}_b^{i}\right).
\end{equation*}

Overall, different receiver designs provide distinct trade-offs between computational complexity, reliance on explicit modeling assumptions, and robustness to impairments. We henceforth adopt a unifying perspective in which any such receiver is viewed as a {\em primary detector} that produces soft outputs, and consider the problem of {\em refining the \acp{llr}} to improve their reliability and scale across modulations, as formulated next. 

\subsection{Problem Formulation}
\label{ssec:Problem} 

In this work, we address the problem of {\em enhancing soft detection} in scenarios where the underlying channel deviates from the assumptions made during the design or training of the receiver. As discussed in the previous subsection, existing detectors (whether model-based or data-driven) produce approximate \acp{llr} whose reliability may significantly degrade under model mismatches (such as \ac{ici}, synchronization errors, or hardware-induced nonlinearities), and may be tailored to a given modulation. 


Our goal is to develop a {\em neural augmentation module} that operates on top of a given receiver (the {\em primary detector}) to refine and possibly rescale its soft outputs. 
Specifically, let $N$ denote the maximum number of bits per symbol supported by the augmentation framework, and $N' \leq N$ the number of bits per symbol for which the primary detector provides \ac{llr} estimates. Accordingly, the primary detector is represented by the mapping $\psi$, which produces the estimated \acp{llr} written as
\begin{equation}
    \label{eqn:PrimaryDet}
    \hat{\myMat{L}}^i = \psi (\myMat{Y}^i) \in \mathbb{R}^{B \times K \times {N'}},
\end{equation}
where $[\hat{\myMat{L}}^i]_{b,k,n}$ represents the \ac{llr} assigned to the $n$th bit at subcarrier $b$ for user $k$. The  augmentation module should thus refine the primary detector \acp{llr} in \eqref{eqn:PrimaryDet} along with the channel output $\myMat{Y}^i$ into estimates of the \acp{llr} for the full set of $B \cdot K \cdot N_s$ bits that are encoded in $\myMat{S}^i$. Table~\ref{tab:symbolssummary} summarizes the key symbols used throughout the paper. 

\begin{table}
\centering
\caption{\normalfont Summary of Symbols}
\begin{tabular}{ll}
\toprule
{\bf Symbol} & {\bf Description} \\
\midrule
$K$          & Number of users \\
$n_r$        & Number of receive antennas \\
$B$          & Number of subcarriers \\
$\mathcal{S} \subset \mathbb{C}$ & Symbol constellation \\
$N_s$        & Number of bits per symbol ($|\mathcal{S}| = 2^{N_s}$) \\
$N$          & Maximum bits per symbol supported by ESCNN \\
$N'$         & Bits per symbol provided by primary detector ($N' \leq N$) \\
\bottomrule
\end{tabular}
\label{tab:symbolssummary}
\end{table}

To ensure broad applicability and practical relevance, we require the augmentation to satisfy the following key properties:
\begin{enumerate}[label={R\arabic*}]
\item \label{itm:channel} \textbf{Model-agnostic operation:} The augmentation should not rely on a specific parametric channel model, and must generalize across diverse channel conditions and impairments.
\item \label{itm:multialg} \textbf{Architecture-agnostic compatibility:} The augmentation should be applicable to different receiver designs, i.e., $\psi(\cdot)$ in \eqref{eqn:PrimaryDet} can be a model-based, data-driven, or a hybrid receiver, and the augmentation should not require modifications to its internal structure.
\item \label{itm:latency} \textbf{Low-latency implementation:} The augmentation should incur minimal computational overhead, ensuring compatibility with real-time receiver processing constraints.
\item \label{itm:constellation} \textbf{ Modulation scaling / LLR completion:} The augmentation should support settings where the number of bits recovered by the primary detector differs from that encoded in the symbol, i.e.,  
$N' \neq N_s$ in \eqref{eqn:PrimaryDet}.

%
%
\end{enumerate}
The above requirements reflect our key design principles. Requirement~\ref{itm:channel} promotes robustness to modeling inaccuracies by ensuring that the augmentation does not depend on a specific channel formulation, and can thus adapt to a wide range of propagation conditions and hardware impairments. Requirement~\ref{itm:multialg} enforces compatibility with diverse receiver designs, allowing the augmentation to be seamlessly integrated with different primary detectors without redesign, thereby facilitating practical deployment across heterogeneous systems. Requirement~\ref{itm:latency} ensures that the benefits of augmentation do not come at the cost of excessive computational overhead, preserving the real-time operation of the receiver chain.

Requirement~\ref{itm:constellation} further extends the notion of invariance introduced in~\ref{itm:multialg} by decoupling the augmentation from the specific modulation scheme used by the primary detector. In particular, it enables receivers that are designed or tuned for a specific number of bits per symbol (i.e., $N' < N$) to be effectively applied in higher-order modulation settings.  This arises in two distinct scenarios: $(i)$ the primary detector is configured for a lower-order modulation scheme than the architecture supports (modulation scaling); or $(ii)$ the primary detector computes only a subset of the bit LLRs for the current modulation (LLR completion). 
Scenario $(i)$ corresponds to, e.g., using deep learning-based receivers that are trained for a fixed modulation scheme (e.g., QPSK \cite{honkala2021deeprx}), and exhibit limited generalization to other constellations. 
Scenario $(ii)$ arises, for example, in model-based detectors such as sphere decoders, whose computational complexity grows rapidly with the constellation size, often limiting their practical use to lower-order modulations. 
Requirement \ref{itm:constellation} indicates that the augmentation should enable these receivers to be extended to higher-order modulations without altering the primary model, thereby enhancing their flexibility and scalability.

To meet requirements~\ref{itm:channel}--\ref{itm:constellation}, we adopt a data-driven design approach. Specifically, during the training phase, we assume access to the primary detector as well as to a dataset of channel outputs and their corresponding transmitted bits (which can vary between samples), given by  
\begin{equation}
    \label{eqn:dataset}
    \mySet{D} = \left\{\myMat{Y}^j, \myMat{D}^j, N_s^j \right\}_{j=1}^{|\mySet{D}|}.
\end{equation}
where $N_s^j \in \{1, \ldots, N\}$ denotes the number of transmitted bits of the $j$-th training sample, i.e., the number of bits per symbol encoded in $D^j$. This dataset is used to learn a mapping that refines the \acp{llr} produced by the primary detector, enabling improved performance under realistic, potentially mismatched operating conditions.

\section{Receiver Neural Augmentation}
\label{sec: Method}

In this section, we present the proposed neural augmentation framework for soft detection. The design of the augmentation is directly guided by the requirements outlined in Subsection~\ref{ssec:Problem}, namely: channel-model-agnostic operation (\ref{itm:channel}), compatibility with diverse receiver architectures (\ref{itm:multialg}), low-latency implementation (\ref{itm:latency}), and  supporting  modulation scaling (\ref{itm:constellation}).

To satisfy \ref{itm:channel},   we adopt a data-driven approach in which a \ac{dnn} is trained to refine the primary detector. Rather than relying on an explicit parametric channel model, the augmentation learns to correct the soft outputs directly from data. Our design is inspired by the DeepSIC algorithm~\cite{DeepSIC2021}, which demonstrated the effectiveness of learning interference mitigation in complex, nonlinear settings. Extending this principle to multi-carrier systems, the proposed augmentation is tailored to capture and compensate for cross-subcarrier dependencies, enabling effective mitigation of \ac{ici} and other non-ideal effects arising in practical \ac{mimo}-\ac{ofdm} channels.

To address both \ref{itm:multialg} and \ref{itm:constellation}, the augmentation is designed to operate on top of the {\em soft outputs} of the primary detector, namely, its bit-wise \acp{llr} in \eqref{eqn:PrimaryDet}. This choice provides a unified interface that is agnostic to the internal structure of the receiver, allowing seamless integration with model-based, data-driven, and hybrid architectures. Moreover, since \acp{llr} represent bit-wise probabilities, they naturally decouple the augmentation from the specific modulation scheme used by the primary detector, enabling compensation for modulation mismatches and supporting scalable receiver implementations. 

To meet \ref{itm:latency}, we design the augmentation to be lightweight and efficient. The proposed architecture, termed \ac{ourdnn}, is based on convolutional layers with kernels spanning adjacent subcarriers, allowing it to exploit local spectral correlations while maintaining low computational complexity. The use of shared convolutional kernels promotes parameter efficiency by leveraging approximate shift-invariance across subcarriers, while element-wise scaling enables adaptation to subcarrier-specific variations. Furthermore, the architecture is structured as a set of parallel per-user modules, facilitating efficient hardware implementation and enabling improved generalization and faster training with limited data~\cite{uzlaner2025async}.

We next present the proposed method. Subsection~\ref{subsec: Architecture} introduces the augmentation architecture. The training procedure is described in Subsection~\ref{subsec: Training}, and a discussion of the resulting properties is provided in Subsection~\ref{subsec: Discussion}.

\subsection{Architecture}
\label{subsec: Architecture} 
The main building block of our augmentation architecture is the \emph{\ac{ourdnn}}. Each \ac{ourdnn} module is dedicated to a single user, and its task is to map the soft information produced by the primary detector, denoted $\hat{\myMat{L}}^i \in \mathbb{R}^{B\times K \times N}$, along with the corresponding channel outputs ${\myMat{Y}}^i \in \mathbb{C}^{B \times n_r}$, into enhanced \acp{llr}, which we denote by  $\tilde{\myMat{L}}^i_k\in \mathbb{R}^{B \times N}$ for the $k$th user. The operation  
is comprised of two key stages: $(i)$ feature pre-processing and $(ii)$ learned processing that applies the \ac{ourdnn} modules to the extracted features..  

\subsubsection{Feature Pre-Processing}
To support modulation scaling and LLR completion (\ref{itm:constellation}), when the primary detector produces $N'<N$ \acp{llr} per symbol, the missing entries are zero-filled prior to further processing, where a zero \ac{llr} corresponds to an uninformative soft output with equal probability for both bit values. The zero-padded primary detector \acp{llr} are given by the tensor $\hat{\myMat{L}}_{\rm ext}^i \in \mathbb{R}^{B \times K \times N}$, whose elements are given by 
\begin{equation}
    \big[\hat{\myMat{L}}_{\rm ext}^i \big]_{b,k,n}= \begin{cases}
        \big[\hat{\myMat{L}}^i \big]_{b,k,n} & n \leq N' \\
        0 & n> N'
    \end{cases}.
    \label{eqn:ZeroFill}
\end{equation}

For numerical stability, we then convert the (possibly zero-padded) input \acp{llr} into probabilistic values bounded in $[0,1]$. This conversion into probabilities uses the sigmoid function, denoted $\sigma(\cdot)$, leveraging the observation that
\begin{align}
    \sigma(l_{b,k,n}^i) &= \frac{1}{1+\exp(-l_{b,k,n}^i)}
    \stackrel{(a)}{=} \frac{1}{1+\frac{p(d^i_{b,k,n} = 0| \myMat{Y}^i)}{p(d^i_{b,k,n} = 1| \myMat{Y}^i)}} \notag \\
     &\stackrel{(b)}{=} p(d^i_{b,k,n} = 1| \myMat{Y}^i),
\end{align}
where $(a)$ stems from \eqref{eqn:LLR}, while $(b)$ follows from the fact that $p(d^i_{b,k,n} = 0| \myMat{Y}^i) = 1-p(d^i_{b,k,n} = 1| \myMat{Y}^i)$. 

Accordingly, the probabilistic features $\sigma(\hat{\myMat{L}}_{\rm ext}^i)$ are reshaped as the $B\times K\cdot N$ matrix $\hat{\myMat{P}}$, and are stacked along with the real and imaginary parts of ${\myMat{Y}}^i$ to form the input matrix $\myMat{X}^i \in \mathbb{R}^{B \times (2n_r + K\cdot N)}$. This representation provides the network with both soft information from the primary receiver about the transmitted bits from all users and the received signal context, while enhancing numerically stable and training-compatible feature values.


\begin{figure} 
    \centering
    \includegraphics[width=\linewidth]{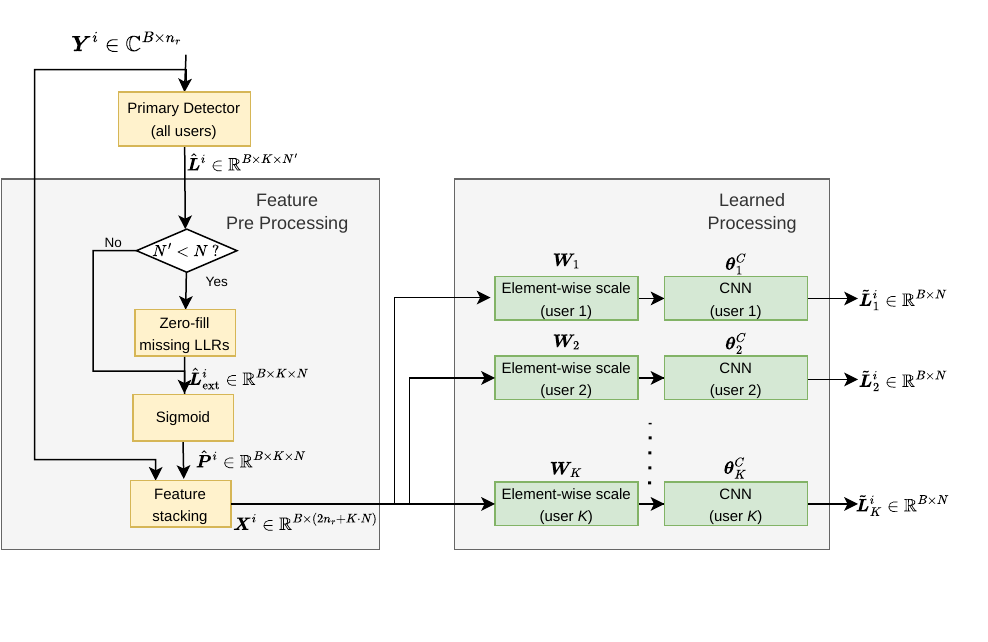}
    \vspace{-1.2cm}
    \caption{\acs{ourdnn} based receiver augmentation architecture at time \textit{i} }
    \label{fig:Augment}
\end{figure}

\subsubsection{Learned Processing}
The structured input $\myMat{X}^i$ is  processed by the learned processing module, which is implemented as an element-wise scale followed by a \ac{cnn} representing a one-dimensional convolutional kernel spanning adjacent subcarriers for each user $k$. The trainable parameters $\myVec{\theta}_k$ include both the   scale factors matrix, denoted $ \myMat{W}_k \in \mathbb{R}^{B\times(2n_r+K \cdot N)}$, and the \ac{cnn} kernels, denoted $\myVec{\theta}_k^{\rm C}$, i.e., $\myVec{\theta}_k = \{\myMat{W}_k, \myVec{\theta}_k^{\rm C}\}$. 
The \ac{cnn}, which is particularly designed to be compact (e.g., in our numerical study we use merely three convolutional layers), produces an output with exactly $N$ channels, corresponding to the maximal number of bits per constellation symbol. The \ac{cnn}, whose mapping is denoted by ${\rm CNN}_{\myVec{\theta}_k^{\rm C}}$, yields bit-wise LLR values which we write as  
\begin{equation}
 \tilde{\myMat{L}}_k^i = f_{\myVec{\theta}_k}(\myMat{X}^i) :=  {\rm CNN}_{\myVec{\theta}_k^{\rm C}} \left(\myMat{W}_k \odot \myMat{X}^i\right)
 \in \mathbb{R}^{B \times N},   
 \label{eqn:output} 
\end{equation}
where $\odot$ is the Hadamard product. This design enables the network to exploit local correlation across neighboring subcarriers, while emphasizing the uniqueness of each subcarrier, and jointly operating over all receive antennas and user probabilities. Note that while the CNN has $N$ supported outputs, only the $N_s$ outputs corresponding to the current modulation are utilized for \acp{llr}.

\subsubsection{Overall Algorithm} 
\label{subsec:overall}

Since each \ac{ourdnn} module is dedicated to refining the \acp{llr} of a single user, we employ $K$ such modules in parallel, one per user in the system. Each module processes the soft outputs of the primary detector corresponding to all users, together with the shared channel outputs, and produces enhanced bit-wise \acp{llr}. The final augmented receiver output is then obtained by concatenating the outputs of all $K$ modules, yielding the complete set of enhanced \acp{llr} as 
\begin{equation}
    [\tilde{\myMat{L}}^i]_{:,k,:} = \tilde{\myMat{L}}_k^i. 
    \label{eqn:OutLLRS} 
\end{equation}
This parallel modular structure 
preserves low latency and supports efficient parallelization in practical implementations. 
The overall procedure is summarized as Algorithm~\ref{alg:Augment}, and illustrated in Fig.~\ref{fig:Augment}.

\begin{algorithm}
\caption{Augmented Receiver at Time $i$}

\label{alg:Augment} 
\SetKwInOut{Init}{Init}
\Init{Primary detector $\psi$;  \ac{ourdnn} weights $\{\myVec{\theta}_k\}_{k=1}^K$}
\SetKwInOut{Input}{Input}
\Input{$\myMat{Y}^i \in \mathbb{C}^{B \times n_r}$}
Apply primary detector $\hat{\myMat{L}}^i = \psi(\myMat{Y}^i) \in \mathbb{R}^{B \times K \times N'}$\;
\If{$N' < N$ \textnormal{(i.e., LLRs are missing)}}{
    Zero-pad $\hat{\myMat{L}}^i \mapsto \hat{\myMat{L}}_{\rm ext}^i \in \mathbb{R}^{B \times K \times N}$ via \eqref{eqn:ZeroFill}\;
}
Convert to bit-wise probabilities $\hat{\myMat{P}}^i = \sigma\big(\hat{\myMat{L}}_{\rm ext}^i\big)$\;
Rearrange $[\hat{\myMat{P}}^i, \myMat{Y}^i] \mapsto \myMat{X}^i \in \mathbb{R}^{B \times (2n_r + K \cdot N)}$\;
\For{$k = 1, \ldots, K$}{
    Apply $k$th \ac{ourdnn}: $\tilde{\myMat{L}}_k^i = f_{\myVec{\theta}_k}\big(\myMat{X}^i\big)$\;
}
\KwRet{$\tilde{\myMat{L}}^i$ as in \eqref{eqn:OutLLRS}}
\end{algorithm}

\subsection{Training Methodology}
\label{subsec: Training}

The goal of the training procedure is to tune the proposed augmentation to operate on top of a given primary detector, tailored for a prescribed modulation scheme with $N'$ bits per symbol. In particular, the augmentation is trained to refine the soft outputs of the primary detector such that the resulting \acp{llr} are reliable and well calibrated for subsequent channel decoding under realistic channel conditions.

To this end, we adopt a data-driven approach in which the augmentation parameters are learned from a dataset $\mySet{D}$ defined in~\eqref{eqn:dataset}. Recall that the feature pre-processing stage maps the primary detector outputs into probabilistic values via the sigmoid function, yielding the structured input $\myMat{X}^i$, which combines soft information from all users with the corresponding channel observations. Each \ac{ourdnn} module, parameterized by $\myVec{\theta}_k$, processes $\myMat{X}^i$ to produce refined probabilistic estimates of the transmitted bits.

\subsubsection*{Loss Function}
The training objective is formulated as the empirical risk associated with predicting the true transmitted bits from the dataset. Specifically, we adopt the binary cross-entropy loss, which is well suited for probabilistic outputs and promotes accurate bit-wise predictions. The empirical loss for the $k$th user is given by 
\begin{align}
\mathcal{L}_{\mySet{D}}(\myVec{\theta}_k) &=  
 -\frac{1}{|\mySet{D}|} \sum_{j=1}^{|\mySet{D}|} \sum_{b=1}^B\frac{1}{{N_s^j}} \sum_{n=1}^{N_s^j} 
\Bigg( d^j_{b,k,n} \log [f_{\myVec{\theta}_k}(\myMat{X}^j)]_{b,n} \nonumber\\
&\quad + \big(1 -  d^j_{b,k,n}\big)
\log \big(1 -  [f_{\myVec{\theta}_k}(\myMat{X}^j)]_{b,n}\big) \Bigg),
\label{eq:training} 
\end{align}
where $f_{\myVec{\theta}_k}(\cdot)$ denotes the mapping implemented by the $k$th augmentation module stated in \eqref{eqn:output}. Note that the loss is evaluated only over the $N_s^j$ active bit outputs, rather than all 
$N$ supported outputs.
The objective \eqref{eq:training} encourages the network to produce probabilistic outputs that are well aligned with the true bit values, resulting in logits that correspond to reliable \acp{llr}. Moreover, the loss is fully differentiable, enabling efficient end-to-end training using gradient-based optimization.

\subsubsection*{Training Procedure}
The parameters $\{\myVec{\theta}_k\}_{k=1}^K$ are thus learned using conventional \ac{dnn} training methods, i.e., mini-batch \ac{sgd} or its variants. In each iteration, a batch of samples from $\mySet{D}$ is processed through the primary detector and the augmentation, the loss in~\eqref{eq:training} is evaluated, and the parameters are updated via backpropagation. The overall training procedure is summarized in Algorithm~\ref{alg:training}.

\begin{algorithm}
\caption{\ac{sgd} Training of Augmented Receiver}
\label{alg:training}
\SetKwInOut{Initialization}{Init}
\Initialization{Step-size $\mu$; \# batches $Q$; epochs $e_{\rm M}$;\\ initial weights $\{\myVec{\theta}_k\}_{k=1}^K$; primary detector $\psi$}
\SetKwInOut{Input}{Input} 
\Input{Training dataset $\mySet{D}$}
\For{$j=1,\ldots, |\mySet{D}|$}
{   
   Apply primary detector $\hat{\myMat{L}}^j = \psi(\myMat{Y}^j)$\;
   Form $\myMat{X}^j$ via feature pre-processing\;
}
\For{$e=1,\ldots, e_{\rm M}$}
{
    Randomly divide $\mySet{D}$ into $Q$ batches $\{\mySet{D}^{(q)}\}_{q=1}^Q$\;
    \For{each batch $\mySet{D}^{(q)}$}
    {
        \For{$k=1,\ldots, K$}
        {
             Apply \ac{ourdnn} $\myVec{\theta}_k$ to each  $\myMat{X} \in \mySet{D}^{(q)}$\;
     
        Compute loss $\mathcal{L}_{\mySet{D}^{(q)}}(\myVec{\theta}_k)$ using \eqref{eq:training}\;
        Update  $\myVec{\theta}_k \gets \myVec{\theta}_k - \mu\nabla_{\myVec{\theta}_k} \mathcal{L}_{\mySet{D}^{(q)}}(\myVec{\theta}_k)$\;
        }
    }
} 
\KwRet{$\{\myVec{\theta}_k\}_{k=1}^K$}
\end{algorithm}




 \subsection{Discussion}
 \label{subsec: Discussion} 
\subsubsection*{Suitability for Handling Requirements~\ref{itm:channel}--\ref{itm:constellation}}
The proposed augmentation framework is designed to directly address the requirements outlined in Subsection~\ref{ssec:Problem}. In particular, it satisfies \ref{itm:channel} by adopting a data-driven design that does not rely on an explicit channel model, enabling robustness to a wide range of impairments and mismatches. Requirement~\ref{itm:multialg} is met by operating on the soft outputs of the primary detector, allowing seamless integration with diverse receiver architectures without modifying their internal structure. Requirement~\ref{itm:latency} is addressed through the use of a compact and efficient architecture, ensuring that the augmentation introduces only minimal processing overhead.
The framework also satisfies \ref{itm:constellation}, extending the notion of invariance to the modulation domain. Since the augmentation operates on bit-wise \acp{llr}, it is inherently decoupled from the specific constellation assumed by the primary detector.  As a result, receivers designed for lower-order constellations can be effectively applied in higher-order settings, either by enhancing incomplete soft information or by correcting mismatched outputs. 

\subsubsection*{Complexity}
An important aspect of the proposed design is its low computational complexity. As we empirically demonstrate in Section~\ref{sec: Numerical Study}, the latency overhead induced by Algorithm~\ref{alg:Augment} is minor and often negligible compared to that of the primary detector. From a model complexity perspective, the number of trainable parameters remains small due to the use of \acp{cnn} with short one-dimensional kernels and shared weights across subcarriers. 

To quantify its complexity, consider a three-layer \ac{cnn} with $3 \times 1$ kernels, where the number of output channels in the first and second layers are $\alpha_1$ and $\alpha_2$, respectively. In this case, each \ac{ourdnn} module includes $B \cdot (2n_r + K \cdot N)$ parameters for the element-wise scaling, and $3 \cdot \big((2n_r + K \cdot N + \alpha_2)\alpha_1 + N \cdot \alpha_2\big)$ parameters for the convolutional layers. In the settings considered in Section~\ref{sec: Numerical Study}, this results in only a few thousand parameters, amounting to less than $1\%$ of the parameter count of DeepRx applied to the same scenario. 
Combined with the modular per-user structure, this compactness facilitates efficient training with limited data and enables low-latency inference suitable for practical hardware implementations.

\subsubsection*{Universality across operating regimes} A key property of the proposed framework, which extends beyond the impairment-compensation perspective of \cite{eger2026learning}, is its unified treatment of two structurally distinct operating regimes: \ac{llr} completion, where reduced-complexity detection leaves a subset of bit \acp{llr} unestimated, and \ac{llr} corruption due to hardware impairments such as \ac{cfo}, \ac{iqmm}, or clipping. In both cases, the augmentation refines and reconstructs the soft information through the same mechanism, and can simultaneously handle cases where \acp{llr} are both missing and corrupted by impairments. This task-agnostic nature follows directly from the data-driven approach: observing only the soft outputs of the primary detector and the received signal, the augmentation learns to correct both the sign and magnitude of the incoming \acp{llr}, producing well-calibrated outputs suitable for channel decoding without requiring knowledge of the operating regime or the nature of the impairments.

\subsubsection*{Future Extensions}
The proposed augmentation framework admits several natural extensions. First, since Algorithm~\ref{alg:Augment} is designed to refine input \acp{llr}, multiple augmentation modules could be cascaded, forming a deeper architecture reminiscent of learned unfolding methods~\cite{monga2021algorithm}. Further, while the current design emphasizes compactness and offline training, its integration with online learning~\cite{gusakov2025rapid}, modular adaptation~\cite{raviv2025modular}, and drift detection mechanisms~\cite{uzlaner2025async} constitutes a potential direction for future work. 
\section{Numerical Study}
\label{sec: Numerical Study}
In this section we present our experimental study. We first detail the experimental setup in Subsection~\ref{ssec:expSetup}. Then, we present the dedicated studies the corresponding results in Subsection~\ref{ssec:expRes}.

\subsection{Experimental Setup}
\label{ssec:expSetup}
We numerically evaluate Algorithm~\ref{alg:Augment} in several relevant multi-user MIMO scenarios with different representative primary receivers, detailed as follows.

\subsubsection{MIMO-OFDM Channels}
We use two types of channel models. The first is the tapped delay line (TDL) channel, implemented using the NVIDIA Sionna library \cite{sionna2022}, specifically TDL-C with low and medium spatial correlation levels (low correlation, unless stated otherwise). The second is spatially consistent channel models implemented using QuaDRiGa \cite{jaeckel2014quadriga}, specifically the Rural Macro (RMa) and Urban Micro (UMi) models.

For all scenarios we use $B=24$ subcarriers per OFDM symbol with a subcarrier offset of 30kHz, and simulate $K=4$ users, each transmitting symbols modulated using 16QAM or 64QAM constellations. The subcarriers within each set of 14 OFDM symbols convey the encoded transport block, with the CRC included. The receiver is equipped with $n_r\in \{4,8\}$ antennas, and the signals from these antennas are fed into the soft detectors.

\subsubsection{Compared Algorithms}
We compare four primary detectors: two model-based detectors based on LMMSE equalization and sphere decoder, as well two \ac{dnn}-based receivers based on  DeepSIC \cite{DeepSIC2021} and DeepRX~\cite{honkala2021deeprx}. 
For the purpose of performing channel estimation, which is required by the model-based algorithms, we provide the receivers with a signal with the same exact channel, noise  and impairments as the original signal, but with staggered symbols such that symbols from different users do not overlap. Due to the increased complexity of sphere decoding for $K=4$ users in $64$-QAM modulations, which requires search over $64^4$ joint hypotheses, we utilize in such settings a reduced-bit sphere decoder (RBSD), which recovers $N'=4$ bits. This primary receiver thus operates on a reduced tree of size $16^4$ by grouping  the 64-QAM constellation into 16 symbol sets, each uniquely identified by 4 specific bit positions per symbol, corresponding to a setting where $N' \neq N = 6$ (\ref{itm:constellation}).

Among the \ac{dnn}-based receivers, DeepSIC is a narrowband algorithm, thus we run $B$ DeepSIC instances, one for each subcarrier, while DeepRX jointly processes all subcarriers. We run these four primary detectors first as is, and then augmented by our \ac{ourdnn}. 
The number of training samples and epochs used for training each module in the subsequent experiments  are summarized in Table~\ref{tab:training_config}, while our full source code is available online at~\url{https://github.com/oryeger/DeepOFDM/}.

\begin{table}
\centering
\caption{\normalfont Training configurations per experiment}
\resizebox{\columnwidth}{!}{%
\begin{tabular}{|c|c|c|c|c|}
\hline
{\bf Figure} & {\bf Epochs} & {\bf Primary Train} & {\bf ESCNN Train} & {\bf Test} \\ \hline
Figs.~\ref{fig:CLEAN}--\ref{fig:IQMM} & 150 & 2500 & 2000 & 10000 \\ \hline
Fig.~\ref{fig:Completion}              & 150 & 6700 & 5300 & 27700 \\ \hline
Fig.~\ref{fig:CLIP}--\ref{fig:RBSD_CFO} & 500 & 1700 & 1300 & 6700  \\ \hline
\multicolumn{5}{l}{\footnotesize $^*$Primary Train, ESCNN Train, and Test are in QAM symbols.} \\
\end{tabular}
}
\label{tab:training_config}
\vspace{-0.2cm}
\end{table}


\subsubsection{Performance Metrics}
For all tests we generate random bits for each user, add cyclic redundancy code (CRC) and perform channel encoding using Low Density Parity Check (LDPC) code. Then, the bits are modulated and transmitted over a wireless channel, where noise and other impairments are introduced. At the receiver side, the resulting LLRs from the unaugmented and augmented detectors are supplied to a soft LDPC decoder. After decoding, the CRC is verified.

We evaluate performance using two complementary metrics. The bit-wise mutual-information (MI) between the transmitted bits and the LLR outputs is computed using histogram-based density estimation~\cite{fertl2010mi}. In the bit-interleaved coded modulation framework, MI provides a direct measure of LLR quality, reflecting how much information the LLRs carry about the transmitted bits, both before and after refinement, independently of any specific code rate, code block length, or decoder implementation. We additionally report the block error rate (BLER) for a representative code rate, computed based on the CRC pass/fail outcome, which provides a concrete demonstration of the end-to-end gains in a practical receiver chain.


 \subsection{Results}
 \label{ssec:expRes}
 We next present the results of our experimental study evaluating the proposed \ac{ourdnn}-based augmentation. 
Our study gradually evaluates different capabilities of our proposed methodology in: 
$(i)$ refining \acp{llr} under hardware impairments; 
$(ii)$ being transferable across different primary receivers and different modulations; 
$(iii)$ completing \acp{llr} for primary receivers that compute only a subset of the bit LLRs; 
and $(iv)$ inducing limited excessive latency.
 

\subsubsection{\ac{llr} Refinement}
\begin{figure}
    \centering
    \includegraphics[width=\columnwidth]{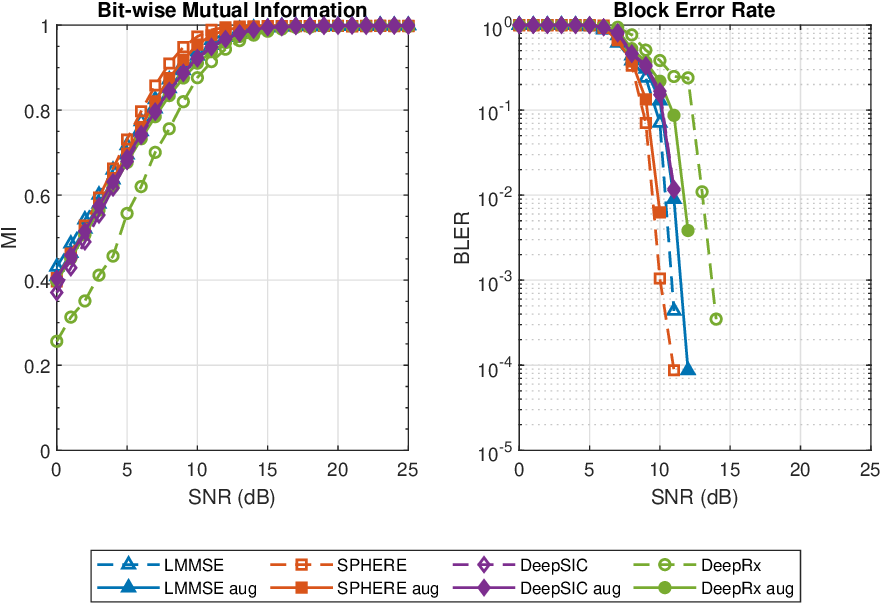}     
    \caption{MI and BLER (LDPC code-rate $0.82$) versus SNR, TDL-C channel, 16QAM modulation, 4 users, $n_r=8$}
    \label{fig:CLEAN}
    \vspace{-0.1cm}
\end{figure}

We first evaluate the ability of ESCNN to refine LLR estimates in the presence of hardware impairments for receivers with $n_r=8$ antennas. We evaluate the performance of 16QAM modulation over a TDL-C channel with a delay spread of 300 ns. In all cases, zero Doppler and low spatial correlation are assumed. Three different 16QAM scenarios are evaluated. In the first scenario, whose results are reported in Fig.~\ref{fig:CLEAN}, we evaluate the baseline case with a linear channel and no ICI. As expected, all detectors achieve similar bit-wise MI and BLER performance regardless of augmentation, since this is the setting all detectors are designed or trained for. Accordingly, the augmentation offers little gain in this case. The exception is DeepRx, which due to its larger model capacity requires more training data and epochs to converge, and the augmentation helps stabilize its performance.

\begin{figure}
    \centering
    \includegraphics[width=\columnwidth]{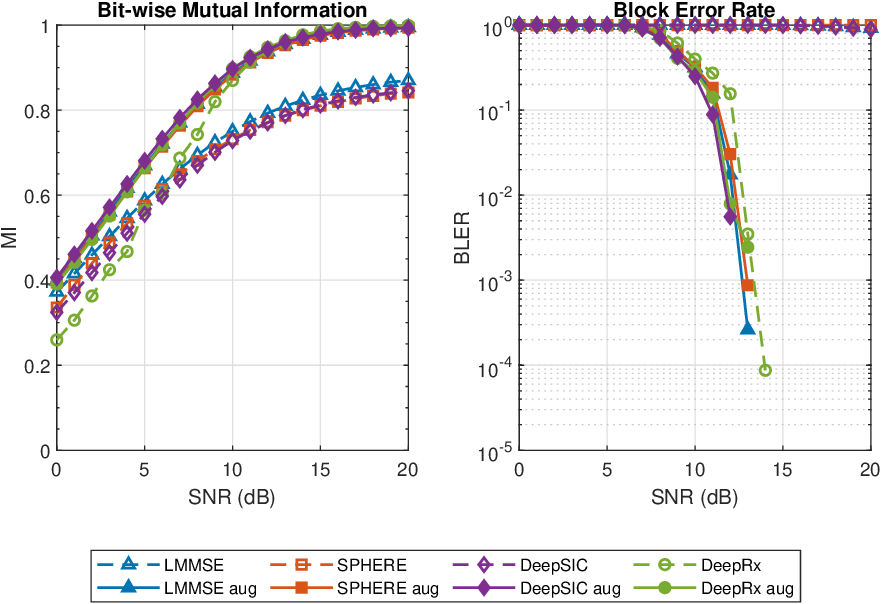}   
    \caption{MI and BLER (LDPC code-rate $0.82$) versus SNR, TDL-C channel, 16QAM modulation, 4 users, $n_r=8$, CFO: 4.5 kHz}
    \label{fig:CFO}
    \vspace{-0.1cm}
\end{figure}

 In the second scenario shown in Fig.~\ref{fig:CFO} we introduce \ac{cfo} of 4.5 kHz, which causes ICI. The CFO is applied separately for the subcarriers of each transport block. We observe that the unaugmented detectors which assume that there is no ICI (LMMSE, sphere, and DeepSIC) perform very poorly, and their MI fails to reach unity, reflecting the uncompensated interference. Conversely, the unaugmented DeepRx that jointly processes all subcarriers and effectively handles ICI. When augmenting the malfunctioning LMMSE, sphere and DeepSIC detectors using \ac{ourdnn}, the MI of all three is driven close to unity, their performance improves significantly, and they are able to reach very low BLERs. For DeepRx, we observe that the augmentation neither improves nor degrades performance. 
 
\begin{figure}
    \centering
    \includegraphics[width=\columnwidth]{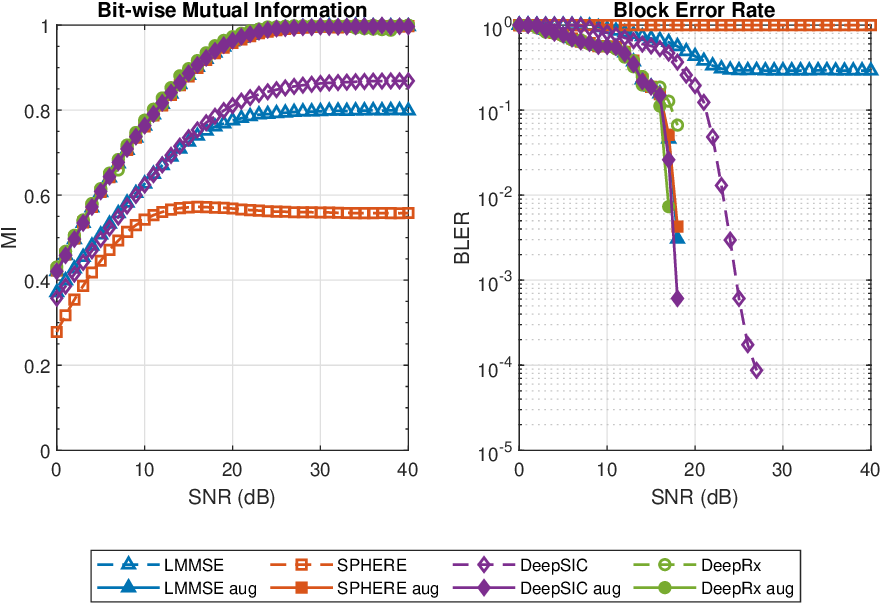}   
    \caption{MI and BLER (LDPC code-rate $0.7$) versus SNR, RMa channel, 16QAM modulation, 4 users, $n_r=8$, CFO: 4.5 kHz}
    \label{fig:RMa}
    \vspace{-0.1cm}
\end{figure}

\begin{figure}
    \centering
    \includegraphics[width=\columnwidth]{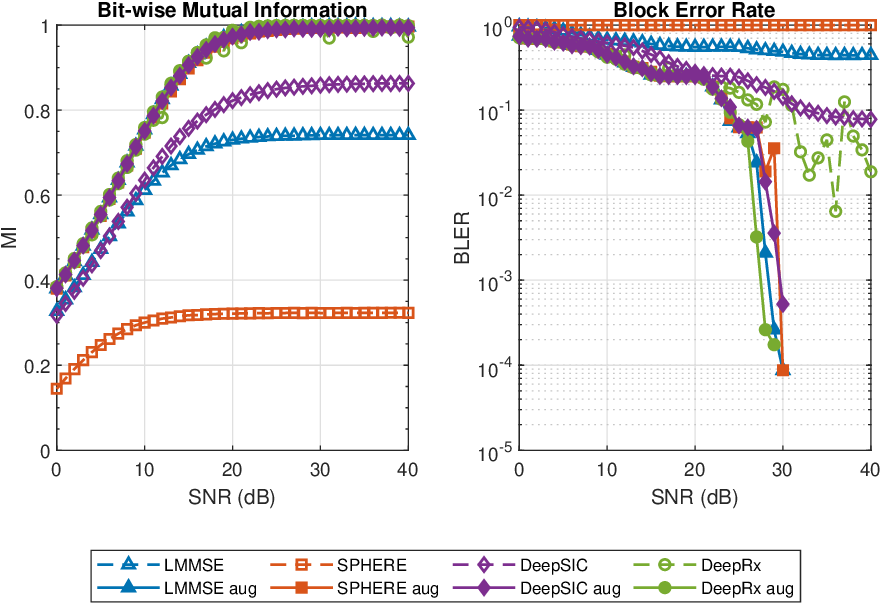}   
    \caption{MI and BLER (LDPC code-rate $0.7$) versus SNR, UMi, 16QAM modulation, 4 users, $n_r=8$, CFO: 4.5 kHz}
    \label{fig:UMi}
    \vspace{-0.1cm}
\end{figure}

 The results observed  in Fig.~\ref{fig:CFO} for TDL-C channels with \ac{cfo} are consistent across different channel models. Specifically, the study is repeated under  Quadriga RMa and  UMi channel models, with the results reported in  Fig.~\ref{fig:RMa} and Fig.~\ref{fig:UMi}, respectively. There, we again consistently show the ability of our proposed augmentation in enabling the struggling primary receiver to cope with impairments induced by \ac{cfo}.

\begin{figure}
    \centering
    \includegraphics[width=\columnwidth]{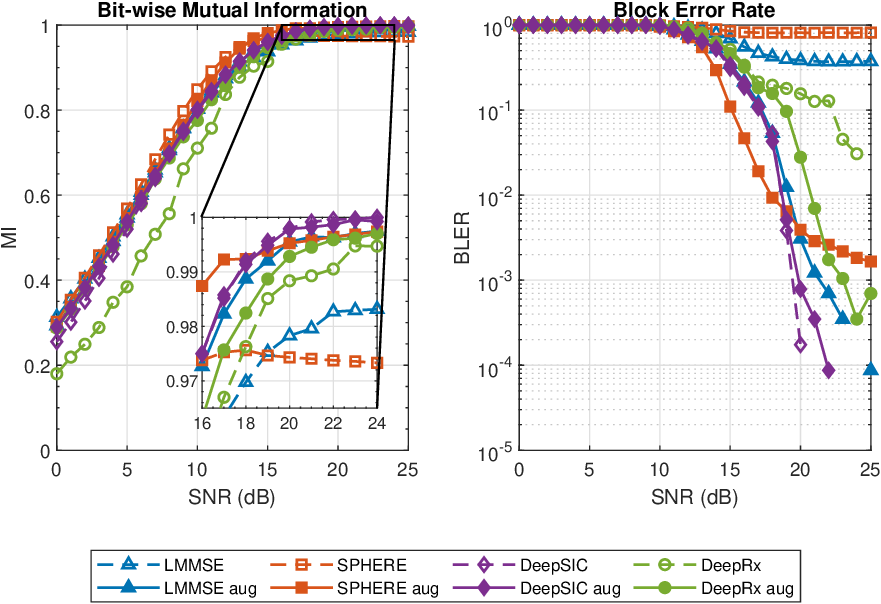}  
    \caption{MI and BLER (LDPC code-rate $0.92$) versus SNR, TDL-C channel, 16QAM modulation, 4 users, $n_r=8$, IQMM: 1 dB, $5^\circ$}    
    \label{fig:IQMM}
    \vspace{-0.1cm}
\end{figure}

 In the next scenario shown in Fig.~\ref{fig:IQMM} we introduce \ac{iqmm}, which causes nonlinear signal distortion. We apply a gain mismatch of $1$ dB and a phase mismatch of $\phi=5^\circ$ over the received signal. 
From Fig.~\ref{fig:IQMM} we observe that the unaugmented LMMSE and sphere decoder struggle as they assume the channel is linear and are unable to achieve reliable decoding. However, the unaugmented DeepSIC does not make the linearity assumption and achieves good results. The proposed ESCNN augmentation drives the MI of all detectors much closer to unity, significantly improving LMMSE and sphere decoder performance. For this scenario, DeepRx  demonstrates stable learning, while the augmentation further improves its BLER performance.

\subsubsection{Transferability}
We next show the transferability of our methodology, i.e., that it can be transferred between primary detectors and between modulations. Table~\ref{tab:detector_transferability} reports the SNR needed to achieve $0.1$ BLER under the same setup as Fig.~\ref{fig:CFO}: a TDL-C channel with a 4.5 kHz \ac{cfo}, code rate 0.82, and 16QAM modulation. Here, we use four different \ac{ourdnn}  modules, each trained using LLRs from one of the four detectors (LMMSE, Sphere, DeepSIC, and DeepRx), while testing is always performed with LLRs from LMMSE. 
The results show that while the unaugmented receiver fails to cope with such ICI (not achieving $0.1$ BLER in any of the considered \acs{snr}s), its augmentation trained with the LMMSE \acp{llr} achieves the best performance as expected. Sphere and DeepSIC augmentations yield only a slight degradation, since they are single-band methods and their augmentation still helps reduce ICI in a general way. For DeepRx, the primary detector itself already compensates for ICI, so the augmentation mainly provides additional support. Although there is still a clear improvement over the unaugmented case, the performance with DeepRx augmentation is noticeably worse compared to the other augmentation strategies.

We proceed to evaluate modulation scaling. To that aim, we evaluate two different ESCNN augmentation DNNs. The first is a dedicated model, whose parameters are specifically designed for the target modulation (i.e., $N = N_s$) and trained only on that modulation. The second is a modulation-agnostic model, whose parameters are configured for the highest modulation order (64QAM) (i.e., $N=6$), and is trained jointly on all three modulations. The dedicated ESCNN is trained using 900 QAM symbols, while the agnostic ESCNN is trained using 2700 symbols divided equally across the three modulation orders (900 per modulation) and randomly shuffled prior to training. Both models are trained for 500 epochs. 

Table~\ref{tab:universal_ESCNN} presents the performance across different modulations under various CFO conditions for the considered \ac{ourdnn} modules with an LMMSE primary receiver. 
The results show that the augmentation performance is not significantly different between the dedicated and agnostic DNNs. Notably, despite being trained across multiple modulation orders, the agnostic model achieves a slight improvement over the dedicated model for 16QAM (8.8 dB vs. 9.5 dB) and 64QAM (17.4 dB vs. 18.1 dB), suggesting that the larger effective training set partially compensates for the generalization constraint.

\begin{table}
\centering
\caption{\normalfont Transferability between primary detectors: measured SNR at 10\% BLER (dB)}
\resizebox{\columnwidth}{!}{%
\begin{tabular}{|c|c|c|c|c|c|}
\hline
Primary detector & \multicolumn{5}{c|}{\bf LMMSE} \\ \hline
Augmentation & {\bf None} & {\bf LMMSE} & {\bf Sphere} & {\bf DeepSIC} & {\bf DeepRx} \\ \hline
SNR (dB) & $\infty$ & 11.7 & 11.8 & 12.1 & 15.3 \\
\hline
\end{tabular}
}
\label{tab:detector_transferability}
\vspace{-0.2cm}
\end{table}

\begin{table}
\centering
\caption{\normalfont Unaugmented vs. dedicated vs. modulation-agnostic ESCNN augmentation: measured SNR at 10\% BLER (dB)}
\resizebox{\columnwidth}{!}{%
\begin{tabular}{|l|c|c|c|}
\hline
Modulation
& QPSK & 16QAM & 64QAM \\
\hline
LDPC code-rate
& 0.59 & 0.7 & 0.75 \\
\hline
CFO (kHz)
& 9 & 4.5 & 2.25 \\
\hline
LMMSE without augmentation
& 5.0 & $14.3$ & $\infty$ \\
\hline
LMMSE with dedicated augmentation
& \textbf{0.6} & \textbf{9.5} & \textbf{18.1} \\
\hline
LMMSE with agnostic augmentation
& \textbf{1.2} & \textbf{8.8} & \textbf{17.4} \\
\hline
\end{tabular}
}
\label{tab:universal_ESCNN}
\vspace{-0.2cm}
\end{table}

\subsubsection{\ac{llr} Completion}
We now investigate the capability of our augmentation to reconstruct missing LLRs. For this scenario we use the 64QAM constellation with a TDL-C channel under {\em medium} spatial correlation, for which sophisticated receiver algorithms such as the sphere decoder typically outperform linear ones. We evaluate the performance of an RBSD  that computes only 4 out of the 6 LLRs per symbol, running an order of magnitude faster than a full sphere decoder at the cost of leaving the remaining 2 bits undetected. To assess this capability, we test a scenario where the number of users equals the number of receive antennas, both set to 4, which is particularly challenging in terms of spatial separation between users.

\begin{figure}
    \centering
        \includegraphics[width=\columnwidth]{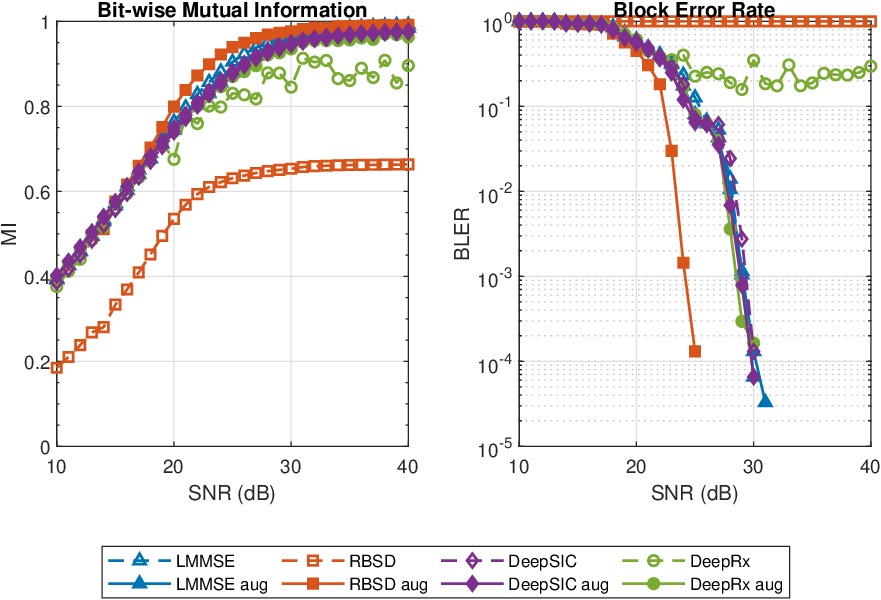}
    \caption{MI and BLER (LDPC code-rate $0.7$) versus SNR, TDL-C channel with medium spatial correlation, 64QAM modulation, 4 users, $n_r=4$}            
    \label{fig:Completion}
    \vspace{-0.1cm}
\end{figure}

In Fig.~\ref{fig:Completion}, the unaugmented RBSD assigns zero to the missing LLRs, which prevents its MI from rising above $\frac{2}{3}$ as expected, resulting in complete BLER saturation. When augmented with ESCNN, the information of the two missing bits is successfully reconstructed, driving the MI to unity and yielding a BLER that outperforms both the unaugmented and augmented LMMSE by 2 to 4 dB, consistent with the inherent advantage of sphere-decoder-based receivers over linear ones. Additionally, ESCNN is able to improve the performance of DeepRx, while DeepSIC achieves performance comparable to LMMSE with or without augmentation.

\begin{figure}
    \centering
    \includegraphics[width=\columnwidth]{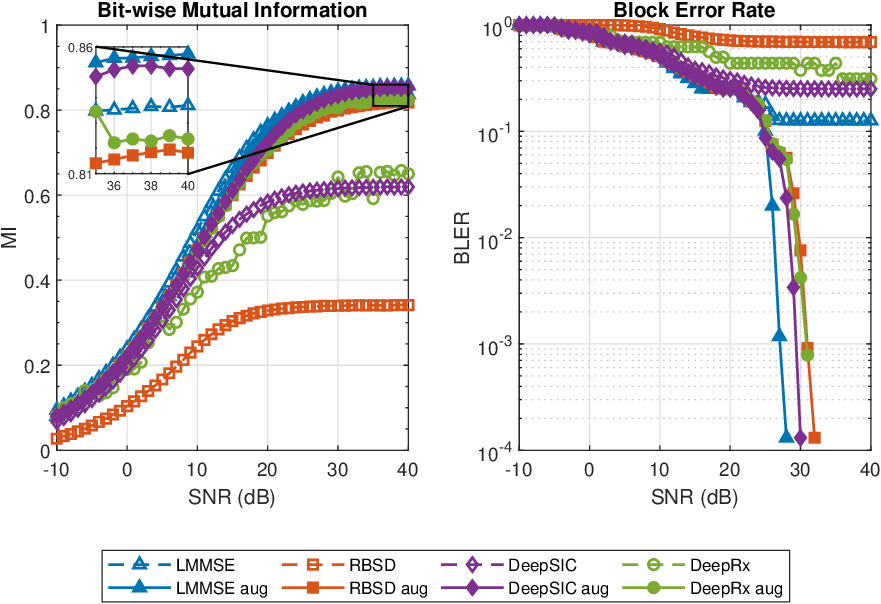}  
    \caption{MI and BLER (LDPC code-rate  $0.4$) versus SNR, UMi channel, 64QAM modulation, 4 users, $n_r=8$, Clip ratio: 3dB}    
    \label{fig:CLIP}
    \vspace{-0.1cm}
\end{figure}

\begin{figure}
    \centering
    \includegraphics[width=\columnwidth]{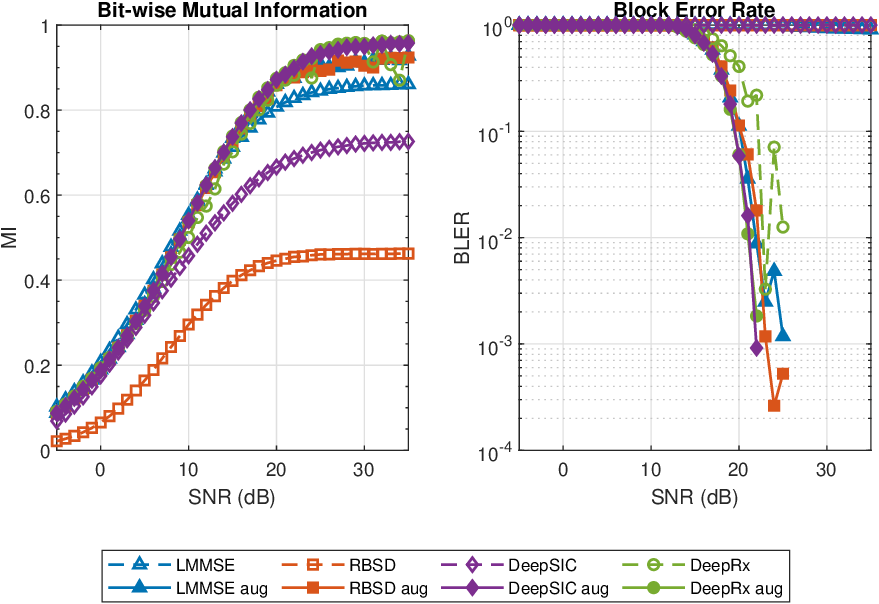}  
    \caption{MI and BLER (LDPC code-rate  $0.75$) versus SNR, TDL-C channel, 64QAM modulation, 4 users, $n_r=8$, CFO: 2.25 kHz}    
    \label{fig:RBSD_CFO}
    \vspace{-0.1cm}
\end{figure}

We further evaluate \ac{ourdnn} in a combined setting that simultaneously exercises both of its core capabilities: LLR reconstruction of the missing bits produced by RBSD, and robustness to hardware-induced distortion with two examples: clipping with ratio of 3~dB , and a CFO of 2.25 kHz. For this  evaluation 64QAM constellation is used, while the clipping impairments is conducted over a UMi channel (Fig.~\ref{fig:CLIP}) and the CFO is conducted over a TDL-C channel (Fig.~\ref{fig:RBSD_CFO}). 

In Fig.~\ref{fig:CLIP} we first note that despite the unaugmented LMMSE exhibiting MI only marginally below its augmented counterpart, its BLER saturates at an error floor with no waterfall behavior. This highlights that MI alone does not fully predict BLER performance, as the latter is also governed by the LDPC code and graph structure, as well as the decoding algorithm employed. We further observe in both Figs.~\ref{fig:CLIP}-\ref{fig:RBSD_CFO} that, across all four detectors, augmentation with ESCNN fully restores the waterfall behavior, yielding BLER curves that decrease smoothly down to less than $10^{-3}$. Under clipping, the augmented receivers consistently achieve an MI above 0.8  at high SNR, though not reaching unity, likely due to the severity of the 3 dB clipping distortion. 

Beyond the common performance gains observed across all detectors, the \ac{ourdnn} augmentation provides an additional benefit for RBSD by reconstructing the two missing LLRs. This substantially increases the MI relative to the unaugmented receiver and enables a clean waterfall region, despite the primary detector operating with only partial soft information. This ability to carry out successful \ac{llr} completion in challenging settings is observed in both Figs.~\ref{fig:CLIP}-\ref{fig:RBSD_CFO}.

\subsubsection{Latency and Complexity}
We conclude our numerical study by assessing the latency overhead induced by the proposed \ac{ourdnn} augmentation. Specifically, to show that our proposed augmentation induces minimal overhead, we report the running time averaged over 5000 OFDM symbols, each comprising 24 subcarriers with 16-QAM modulation, in Table~\ref{tab:running_time}, along with the relative overhead compared to the primary receiver. All timings reported there were measured on the same platform equipped with an NVIDIA RTX 4060 GPU. 
The results in Table~\ref{tab:running_time} show that the overhead introduced by the augmentation is negligible compared to that of either of the primary detectors. Specifically, even for the relatively efficient LMMSE and DeepSIC receivers, the excessive latency is merely $5-6\%$, while being less than $1\%$ of the primary receiver latency for the heavier sphere and DeepRx detectors.  

The low latency overhead of \ac{ourdnn} stems from its lightweight parameterization. To see this, we report in Table~\ref{tab:num_parameters} the overall number of trainable parameters for all \ac{dnn}-aided modules considered in this study. There, we note that the augmentation involves an order of magnitude less parmeters than DeepSIC, and three orders of magnitude less than DeepRx. These values complement the limited wall clock overhead values of Table~\ref{tab:running_time}, showing that our proposed augmentation is indeed relatively lightweight and capable of enhancing existing primary receivers with limited overhead in latency and complexity.

\sisetup{scientific-notation = false}
\begin{table}
\centering
\caption{Measured running time on Nvidia RTX 4060 GPU }
\begin{tabular}{|l|c|c|c|c|}
\hline
 & {\bf LMMSE} & {\bf Sphere} & {\bf DeepSIC} & {\bf DeepRx} \\
\hline
Without augmentation & \SI{70}{\milli\second} & \SI{100}{\second} & \SI{75}{\milli\second} & \SI{800}{\milli\second}\\
\hline
With augmentation & \SI{74}{\milli\second} & $\sim$\SI{100}{\second} & \SI{79}{\milli\second} & \SI{804}{\milli\second}\\
\hline
Relative overhead & 6\% & $\sim$0\% & 5\% & 0.5\%\\
\hline
\end{tabular}
\label{tab:running_time}
\vspace{-0.2cm}
\end{table}
\sisetup{scientific-notation = true}

\begin{table}
\centering
\caption{Number of parameters}
\begin{tabular}{|c|c|c|}
\hline
\textbf{ESCNN} & \textbf{DeepSIC} & \textbf{DeepRx} \\
\hline
\num{80400} & \num{909696} & \num{12377600} \\
\hline
\end{tabular}
\label{tab:num_parameters}
\vspace{-0.2cm}
\end{table}

\section{Conclusion}
\label{sec:conclusion}
We introduced a neural augmentation framework for soft detection in multi-user \ac{mimo}-\ac{ofdm} systems. Rather than replacing existing receiver architectures, the proposed approach operates as a lightweight post-processing module that refines the soft outputs of arbitrary primary detectors, including model-based, data-driven, and hybrid receivers.We developed the \ac{ourdnn} architecture, a compact convolutional network that exploits local spectral dependencies and jointly processes received signals and detector-generated soft information to produce reliable bit-wise \acp{llr}. The resulting framework is model-agnostic, receiver-agnostic, and modulation-aware, enabling a unified treatment of both impairment-induced \ac{llr} corruption and reduced-complexity \ac{llr} completion. Through extensive simulations over realistic 3GPP-compliant channel models and a variety of hardware impairment scenarios, we demonstrated that the proposed augmentation consistently improves the reliability of soft information and significantly enhances decoding performance across diverse receiver architectures, while incurring only negligible computational and latency overhead. These results suggest that neural LLR augmentation constitutes a practical and scalable mechanism for extending the operating range of existing wireless receivers.

\bibliographystyle{IEEEtran}
\bibliography{IEEEabrv,refs}

\end{document}